\documentclass[11pt,twoside,a4paper,cmspaper,final,collab]{cms-tdr}
\def\svnVersion{4df1b52-D}\def\svnDate{2026/08/02}\def\cmsCernNoTag{CERN-EP-2026-090}\def\cmsCernDate{\today}\def\cmsMessage{Submitted to the Journal of High Energy Physics}
\begin{document}\cmsNoteHeader{B2G-24-009}

\newcommand{\runtwolumi}{138\xspace}
\newcommand{\nbjets}{\ensuremath{N_{{\PQb}\text{-jet}}}\xspace}
\newcommand{\ST}{\ensuremath{S_{\mathrm{T}}}\xspace}
\newcommand{\pb} {\mbox{\ensuremath{\,\text{pb}}}\xspace}
\newcommand{\fb} {\mbox{\ensuremath{\,\text{fb}}}\xspace}
\newcommand{\Wh}{\ensuremath{\PW_{\mathrm{h}}}\xspace}
\newcommand{\tWhb}{\ensuremath{\PQt_{\Wh+\PQb}}\xspace}
\newcommand{\mzprec}{\ensuremath{m_{\zp}^{\text{rec}}}\xspace}
\providecommand{\Pj}{\HepParticle{j}{}{}\xspace}

\cmsNoteHeader{B2G-24-009}
\title{Search for a top-philic \texorpdfstring{\zp}{Z'} boson decaying into a \texorpdfstring{\ttbar}{t-tbar} pair in a final state with jets and an electron or muon in proton-proton collisions at \texorpdfstring{$\sqrt{s} = 13.6\TeV$}{sqrt(s) = 13.6 TeV}}

\author[cern]{The CMS Collaboration}

\date{\today}

\abstract{A search for a top-philic \zp boson in a final state with jets and an electron or muon is presented.  The search is based on a sample of proton-proton collision data collected at $\sqrt{s} = 13$\TeV by the CMS experiment at the CERN LHC during 2016--2018, corresponding to an integrated luminosity of $\runtwolumi\fbinv$. The top-philic \zp boson is produced in association with a top-antitop quark pair (\ttbar) and decays into a \ttbar pair, as it couples exclusively to top quarks. The analysis aims to identify a heavy \zp boson that produces Lorentz-boosted top quarks, whose hadronic decay products are merged into large-radius jets. A machine-learning algorithm is employed to identify such jets. The distribution of the invariant mass of the two top quark candidates with the highest transverse momentum is used as the discriminant variable in a \zp boson mass range of 0.5--3\TeV, with intrinsic widths of 4, 10, 20, and 50\% relative to its mass. The results obtained are found to be in agreement with the standard model background prediction. Upper limits at 95\% confidence level are set on the production cross section of the \zp boson, for each of the decay widths as a function of its mass. These results represent the most stringent constraints to date on the existence of a top-philic \zp boson.}

\hypersetup{
pdfauthor={CMS Collaboration},%
pdftitle={Search for a top-philic Z' boson decaying into a ttbar pair in a final state with jets and an electron or muon in proton-proton collisions at sqrt(s) = 13 TeV},
pdfsubject={CMS},%
pdfkeywords={CMS, search for new physics, ttZ', top-philic Z' boson},
}

\maketitle 

\section{Introduction}
\label{sec:Introduction}

The standard model (SM) of particle physics has been remarkably successful in explaining a large number of experimental results. Despite its success, there remain many unknowns, such as the nature of dark matter, the nonzero masses of the neutrinos, an explanation for the origin of matter-antimatter asymmetry in the universe, or the integration of a quantum field theory of gravity. Various theoretical models have been proposed to address these issues, predicting beyond-the-SM (BSM) particles that are primarily produced via the annihilation of a light quark-antiquark pair~\cite{Cullen:2000ef,ref_diquark,Baur:1989kv,ref_axi,Chivukula:2011ng,ref_coloron,Han:2010rf,Leike:1998wr,Zp_SSM_1,Harari:1982xy,Greenberg:1974qb,Pati:1975md,Leonardi:2015qna,lr1,lr3,Dib:2016wge,Cai:2017mow,Gripaios:2009dq,Ajmal:2023yhl,Kramer:2004df,Dorsner:2016wpm}. However, as current exclusion limits on the masses of such new particles extend up to multiple \TeV in mass, alternative hypotheses involving new particles favoring higher-generation fermions, \ie, anogenophilic in nature~\cite{Barbosa:2022mmw}, gain relevance. Here, we examine the extreme scenario of a top-philic resonance that couples only to top quarks, which has garnered substantial attention in the literature within models of dark matter~\cite{Cox:2015afa,Zhang:2012da,Baek:2016lnv,Arina:2016cqj,Buckley:2014fba,DHondt:2015nat}, composite Higgs bosons~\cite{Ferretti:2013kya,Vecchi:2015fma,Agashe:2003zs,Xue:2022kjl,Agashe:2004rs,Xue:2022kjl,Arkani-Hamed:2002ikv,Pomarol:2008bh}, or other scenarios predicting heavy neutral resonances~\cite{Fox:2018ldq,Xue:2013hn,Liu:2015hxi,Abdullah:2019dpu,Leonardi:2018jzn}. The search for a new particle with significant couplings to the top quark is also motivated by the large Yukawa coupling of the top quark to the Higgs boson. This gives rise to a substantial contribution to the quantum corrections of the Higgs boson mass, and requires a BSM mechanism to address the resulting naturalness problem~\cite{Chung:2023iwj,Craig:2022eqo,Giudice:2008bi}.

In this analysis, a simplified model~\cite{Kim:2016plm} involving a color-singlet vector \zp boson that couples solely to top quarks and antiquarks is assumed. The relevant interaction is given by the following renormalizable Lagrangian:
\begin{equation}
\mathcal{L} \supset \PAQt\gamma^\mu (c_\mathrm{L} P_\mathrm{L} + c_\mathrm{R} P_\mathrm{R}) \PQt \zp_\mu = c_\PQt \PAQt\gamma^\mu (\cos \theta P_\mathrm{L} + \sin \theta P_\mathrm{R}) \PQt \zp_\mu .
\end{equation}

\begin{figure}[ht]
\centering
\includegraphics[width=0.55\textwidth]{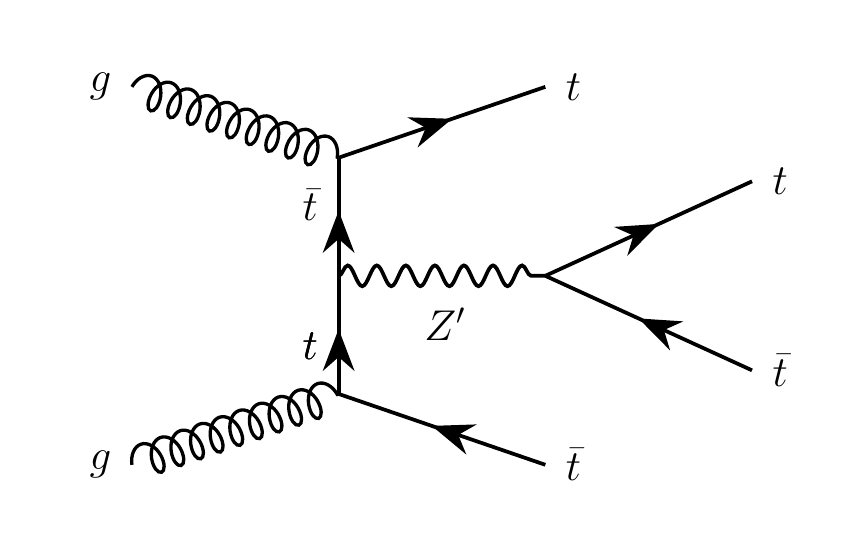}
\caption{Feynman diagram of a top-philic \zp boson produced in association with a top quark pair and decaying to a top quark pair.}
\label{fig:ttZ'_feynman_diagrams}
\end{figure}

Here, $\gamma^\mu$ are the Dirac matrices, $c_\mathrm{L}$ and $c_\mathrm{R}$ are the left- and right-handed couplings of the \zp boson to top quarks, and $c_\PQt = \sqrt{\smash[b]{c_\mathrm{L}^2 + c_\mathrm{R}^2}}$. The variables $P_\mathrm{L}$ and $P_\mathrm{R}$ are the left- and right-handed projection operators, respectively, and the chirality angle $\theta$ is given by $\tan \theta = c_\mathrm{R} / c_\mathrm{L}$. The decay width of the \zp boson relative to its mass can be approximated as $\Gamma_{\zp}/m_{\zp} \approx c_\PQt^2 / (8\pi)$ for a top quark mass lower than the \zp boson mass~\cite{Kim:2016plm}. Both tree- and loop-level production mechanisms are allowed; however, we only consider the former to minimize model dependence. In particular, we focus on the process in which a \zp boson is produced in association with two top quarks, searching for signatures consistent with the decay of four top quarks as shown in Fig.~\ref{fig:ttZ'_feynman_diagrams}. The cross section for this production mechanism is independent of the parameter $\theta$.

In previous analyses by the ATLAS and CMS Collaborations, the first observation of SM four top quark production at $\sqrt{s} = 13$\TeV was reported with observed (expected) significances of 6.1 (4.3) and 5.6 (4.9) standard deviations, respectively~\cite{ATLAS:2023ajo,CMS:2023ftu}. Earlier evidence for this process had been reported by ATLAS and CMS~\cite{ATLAS:2021kqb,CMS:2023zdh}. The ATLAS Collaboration has also conducted searches for a \zp boson produced in association with a pair of top quarks and subsequently decaying via $\zp \to \ttbar$, for \zp boson masses between 1 and 3\TeV in the single lepton final state~\cite{ATLAS:2023taw}. Their corresponding observed (expected) limits on cross-section range from 21 (14)\fb to 119 (86)\fb, depending on the choice of the model parameters. 
In the search reported here, we focus on the single-lepton (electron or muon) channel, where the lepton originates from one of the top quarks, as it has a reasonably small background from quantum chromodynamics (QCD) multijet processes, and can be triggered on with a high efficiency, thereby maintaining a high signal sensitivity. The analysis aims to identify a \zp boson with a mass large enough to produce Lorentz-boosted top quarks, whose decay products may merge into single large-radius jets. We use the machine-learning algorithm \textsc{ParticleNet}~\cite{Qu:2019gqs} for jet identification in this specific topology, for the case of the fully hadronic top quark decay, and we use the invariant mass of the two top quark candidates with the highest transverse momenta as the reconstructed \zp boson mass for the signal extraction.
Additionally, the two associated top quarks provide a powerful means to suppress the SM backgrounds, significantly enhancing the sensitivity of the search.
The search is based on a data sample of proton-proton ($\Pp\Pp$) collisions at a center-of-mass energy of 13\TeV recorded by the CMS experiment
during the years 2016--2018, corresponding to an integrated luminosity of 138\fbinv.

This paper is organized as follows.
Section~\ref{sec:Simulated_signal_and_background_samples} describes the CMS detector and the Monte Carlo (MC) simulated samples for signal and background.
Event selection and event reconstruction are described in Section~\ref{sec:Event_selection}.
Section~\ref{sec:Background_estimation} describes the estimation of the SM background.
Systematic uncertainties are presented in Section~\ref{sec:systematics}.
The results and their interpretation are given in Section~\ref{sec:results}.
The paper is summarized in Section~\ref{sec:summary}.

Tabulated results for this analysis can be found in the HEPData record for this analysis~\cite{hepdata}.

\section{The CMS detector and simulated samples}
\label{sec:Simulated_signal_and_background_samples}

The CMS apparatus~\cite{CMS:2008xjf,CMS:2023gfb} is a multipurpose, nearly hermetic detector, designed to trigger on~\cite{CMS:2020cmk,CMS:2016ngn,CMS:2024aqx} and identify electrons, muons, photons, and (charged and neutral) hadrons~\cite{CMS:2020uim,CMS:2018rym,CMS:2014pgm}. Its central feature is a superconducting solenoid of 6\unit{m} internal diameter, providing a magnetic field of 3.8\unit{T}. Within the solenoid volume are a silicon pixel and strip tracker, a lead tungstate crystal electromagnetic calorimeter (ECAL), and a brass and scintillator hadron calorimeter (HCAL), each composed of a barrel and two endcap sections. Forward calorimeters extend the pseudorapidity coverage provided by the barrel and endcap detectors. Muons are reconstructed using gas-ionization detectors interleaved with the layers of the steel flux-return yoke outside the solenoid. More detailed descriptions of the CMS detector, together with a definition of the coordinate system used and the relevant kinematic variables, can be found in Refs.~\cite{CMS:2008xjf,CMS:2023gfb}. 
Events of interest are selected using a two-tiered trigger system. The first level, composed of custom hardware processors, uses information from the calorimeters and muon detectors to select events at a rate of around 100\unit{kHz} within a fixed latency of 4\mus~\cite{CMS:2020cmk}. The second level, known as the high-level trigger, consists of a farm of processors running a version of the full event reconstruction software optimized for fast processing, and reduces the event rate to a few kHz before data storage~\cite{CMS:2016ngn,CMS:2024aqx}. Triggers requiring the presence of at least one electron or muon, collectively referred to as leptons in the following, are used in our analysis. 

The MC event generators are employed to simulate SM background processes and the signal. The proton structure is described by the parton distribution functions at next-to-next-to-leading order in perturbative QCD~\cite{NNPDF:2017mvq}. The parton shower and hadronization are simulated with \PYTHIA v8.230~\cite{Sjostrand:2014zea}, where the parameters for the underlying event description correspond to the CP5 tune~\cite{Sirunyan:2019dfx}. 

Signal samples are generated at leading order (LO), with a chirality parameter of $\theta = \pi / 4$,
using \MGvATNLO v2.6.5 ~\cite{Alwall:2014hca,Degrande:2014vpa,Branco:1999fs,Degrande:2011ua}, which relies on the model described in Ref.~\cite{Kim:2016plm}. 
For $\theta = \pi/4$, the \zp resonance has no axial component and the quark-initiated $\zp+\Pj$ process at loop level is forbidden~\cite{Greiner:2014qna}.
We focus on the associated production of a \zp boson and top quark pairs, assuming a branching fraction of 100\% for the \zp decay to \ttbar. The \zp boson mass studied ranges from 0.5 to 3\TeV, with samples produced in increments of 250 or 500\GeV for a \zp boson in the range 0.5--2 and 2--3\TeV, respectively. We generate different sets of samples for $\Gamma_{\zp}$ = 4, 10, 20, and 50\% of the \zp boson mass.

The main background in this search is the SM production of a pair of top quarks.
This process is simulated at next-to-LO (NLO) using \POWHEG v2 ~\cite{Nason:2004rx,Frixione:2007vw,Alioli:2010xd,Jezo:2015aia}, and the inclusive cross section is normalized to a next-to-NLO QCD calculation, including soft-gluon terms~\cite{bib:ttbarxs:1,bib:ttbarxs:2,bib:ttbarxs:3,bib:ttbarxs:4,bib:ttbarxs:5,bib:ttbarxs:6,bib:ttbarxs:7}.

Background samples for \ttbar production in association with a boson ($\ttbar\PW$, $\ttbar\PZ$, $\ttbar\PH$, and $\ttbar\PGg$), single top quark production in the $s$ channel or in association with a \PZ boson and jets, and $\PW\PW$, $\PW\PZ$, $\PZ\PZ$, $\PZ\PGg$, $\PW\PH$, and $\PZ\PH$ diboson production are simulated at NLO in QCD with \MGvATNLO. An LO simulation with \MGvATNLO is performed for \ttbar production in association with two bosons, three top quark production, and production of a H in association with a top quark. The \POWHEG generator is used for the single top quark production via the $t$ channel and in association with a \PW boson. These backgrounds are collectively referred to as ``other top'' background in the text. The production of four top quarks is simulated at NLO in QCD with the \MGvATNLO v2.6.5 program. Minor backgrounds, including the production of a photon or a \PZ, \PW, or \PH boson in association with jets, are generated using \PYTHIA and referred to as ``nontop'' processes. 

The event generators are interfaced with a detailed detector simulation based on \GEANTfour~\cite{Agostinelli:2002hh}. Multiple $\Pp\Pp$ interactions within the same or nearby bunch crossings (pileup) are superimposed on the hard scattering process and their frequency distribution is matched to that observed in data. For comparison with observed distributions, the events in the simulated samples are normalized to the same integrated luminosity as the data sample based on their predicted cross sections. Correction factors are applied to the simulation where necessary to improve the description of the reconstruction and identification efficiencies of the final-state objects. 

\section{Event selection and object reconstruction}
\label{sec:Event_selection}

Events are required to match a signature consistent with the decay of four top quarks, with one decaying leptonically and the others hadronically, resulting in signal events characterized by a high level of hadronic activity. A global ``particle-flow'' algorithm~\cite{CMS:2017yfk} is used to reconstruct individual particles in an event, by combining information provided by the tracker, the ECAL, the HCAL, and the muon systems. The reconstructed particles are used to build tau leptons, jets, and missing transverse momentum~\cite{CMS:2018jrd,CMS:2016lmd,CMS:2019ctu}.

Electrons are identified with a multivariate discriminant, which is constructed using variables related to the bremsstrahlung along the electron trajectory, ECAL energy measurements, electromagnetic shower shapes, missing pixel detector hits, and the photon conversion vertex fit probability~\cite{CMS:2020uim}.
A working point with an identification efficiency of 80 (90)\% is chosen for tight (loose) electrons with a transverse momentum $\pt > 33$\,(15)\GeV within a pseudorapidity range $\abs{\eta} < 2.5$ to identify the primary electron candidate (veto additional ones). The momentum resolution for electrons with $\pt \approx 45\GeV$ from $\PZ \to \Pe\Pe$ decays ranges from 1.6 to 5\%. It is generally better in the barrel region than in the endcaps, and also depends on the bremsstrahlung energy emitted by the electron as it traverses the material in front of the ECAL~\cite{CMS:2020uim,CMS-DP-2020-021}. 

Muons~\cite{CMS:2018rym} are measured with detection planes made using three technologies: drift tubes, cathode strip chambers, and resistive-plate chambers. ``Tight''  muons are required to have $\pt > 26$\GeV\,(29\GeV for 2017) and $\abs{\eta} < 2.4$ for the event selection. Additional muons with $\pt > 15$\GeV and $\abs{\eta} < 2.4$ satisfying the ``loose'' requirement are vetoed. The efficiency to reconstruct and identify muons is greater than 96\%. Matching muons to tracks measured in the silicon tracker results in a relative transverse momentum resolution, for muons with \pt up to 100\GeV, of 1\% in the barrel and 3\% in the endcaps. The \pt resolution in the barrel is better than 7\% for muons with \pt up to 1\TeV~\cite{CMS:2018rym}.

All lepton candidates have to pass isolation requirements as defined in Refs.~\cite{CMS:2020uim,CMS:2018rym}, with an efficiency of 95\%.
Selected events must pass a single-lepton trigger and contain exactly one tight electron or muon with zero additional loose candidates.

For each event, hadronic jets are clustered from reconstructed particles using the infrared- and collinear-safe anti-\kt algorithm~\cite{Cacciari:2008gp, Cacciari:2011ma} with a distance parameter of 0.4 and 0.8 for small- and large-radius jets, respectively. Jet momentum is determined as the vectorial sum of all particle momenta in the jet, and is found from simulation to be, on average, within 5 to 10\% of the true momentum over the entire \pt spectrum and detector acceptance. The jet energy resolution amounts typically to 15--20\% at 30\GeV, 10\% at 100\GeV, and 5\% at 1\TeV~\cite{CMS:2016lmd}. Additional selection criteria are applied to each jet to remove jets potentially dominated by anomalous contributions from various subdetector components or reconstruction failures~\cite{CMS:2020ebo}. In this analysis, small-radius jets are required to have $\pt>30\GeV$ and $\abs{\eta}<2.5$.

The missing transverse momentum vector \ptvecmiss is computed as the negative vector sum of the transverse momenta of all the particle-flow candidates in the event, and its magnitude is denoted as \ptmiss~\cite{CMS:2019ctu}. The \ptvecmiss is modified to account for corrections to the energy scale of the reconstructed jets in the event. In this analysis, \ptmiss is required to exceed 60\GeV.

The identification of small-radius jets originating from the decay of \PQb hadrons relies on the DeepJet algorithm~\cite{DeepJet}, which uses information from the calorimeters and tracking detectors~\cite{Sirunyan2018}. In this analysis, we require at least one b-tagged jet using an identification condition with an efficiency of 91.5\%, which corresponds to the loose working point defined in Ref.~\cite{DeepJet}. Corrections are applied to the simulation to match the efficiency measured in data. 

Large-radius jets originating from the hadronic decay of top quarks (\tWhb, \Wh indicating a \PW boson decaying into two quarks) are identified with a mistagging rate of 0.5\% using the machine-learning algorithm \textsc{ParticleNet}~\cite{Qu:2019gqs}, which relies on jet substructure variables, jet constituents, and secondary vertices as inputs to extract information on the particles producing the jets. Events are considered if they have at least two \tWhb candidates with $\pt > 300$\GeV and $\abs{\eta} < 2.4$. The use of a machine-learning algorithm enhances the detection of hadronically decaying boosted top quarks, offering improved discrimination power against the background, composed mainly of SM events with top quarks. Correction factors for this algorithm are applied to the simulation to match the efficiency measured in data~\cite{CMS-PAS-JME-18-002}.

As the signal contains a large number of high-\pt jets, each event is required to have at least six jets (three from the hadronic and one from the leptonic top quark decay, and two \tWhb candidates) and $\HT > 700$\GeV, where \HT is the scalar sum of the \pt of all small-radius jets.

All selected leptons must be separated from small-radius jets by $\Delta R > 0.4$ and from large-radius jets by $\Delta R > 0.8$, where $\Delta R = \sqrt{\smash[b]{(\Delta\eta)^2+(\Delta\phi)^2}}$. Small-radius jets must be separated from large-radius jets by $\Delta R > 1.2$.
This condition ensures that large- and small-radius jets do not overlap and thus justifies the requirement of at least one b-tagged jet in the final state, given the presence of at least two $\tWhb$ candidates. The above criteria are similar to those used in a CMS measurement of the SM production of four top quarks~\cite{CMS:2023zdh}, except that we include more stringent requirements on the number of jets and on \HT. 

Events that pass the above selection criteria enter the signal region (SR) used to search for a top-philic \zp boson.
Using simulated signal samples, we find that the total selection efficiency varies from about 0.26 (0.35) to 4.2 (2.0)\% for \zp bosons with masses between 0.5 and 3\TeV and $\Gamma_{\zp}/m_{\zp} =4$ (50)\%. This efficiency accounts for the event selection and the branching fraction of the decay of four top quarks into a final state with one lepton and jets. We reconstruct a \zp candidate in selected events from the two $\tWhb$ candidates with the highest \pt. The distribution of the invariant mass \mzprec of the \zp candidate is used to search for a BSM signal.

\section{Background estimation}
\label{sec:Background_estimation}

The background in the SR is dominated by events with top quarks from various processes. The $\ttbar+\text{jets}$ process constitutes about 75\% of the total, whereas $\ttbar + \PV$ and single top quark processes ($\PQt+\PV/\PV\PV$, $\PQt\PZ\PQq$) contribute approximately 20\% and the $\ttbar \ttbar$ process accounts for about 4\%. The nontop backgrounds contribute less than 1\%.

To estimate the background expected from the SM, we define a signal-depleted control region (CR) to validate the background modeling in the SR and to constrain the normalization of the total background in a simultaneous maximum likelihood fit. This region is obtained by requiring events contain zero b-tagged jets, while keeping all other selections the same as described in Section~\ref{sec:Event_selection}, making it orthogonal to the SR.

\begin{figure}[ht]
\centering
\includegraphics[width=0.45\textwidth]{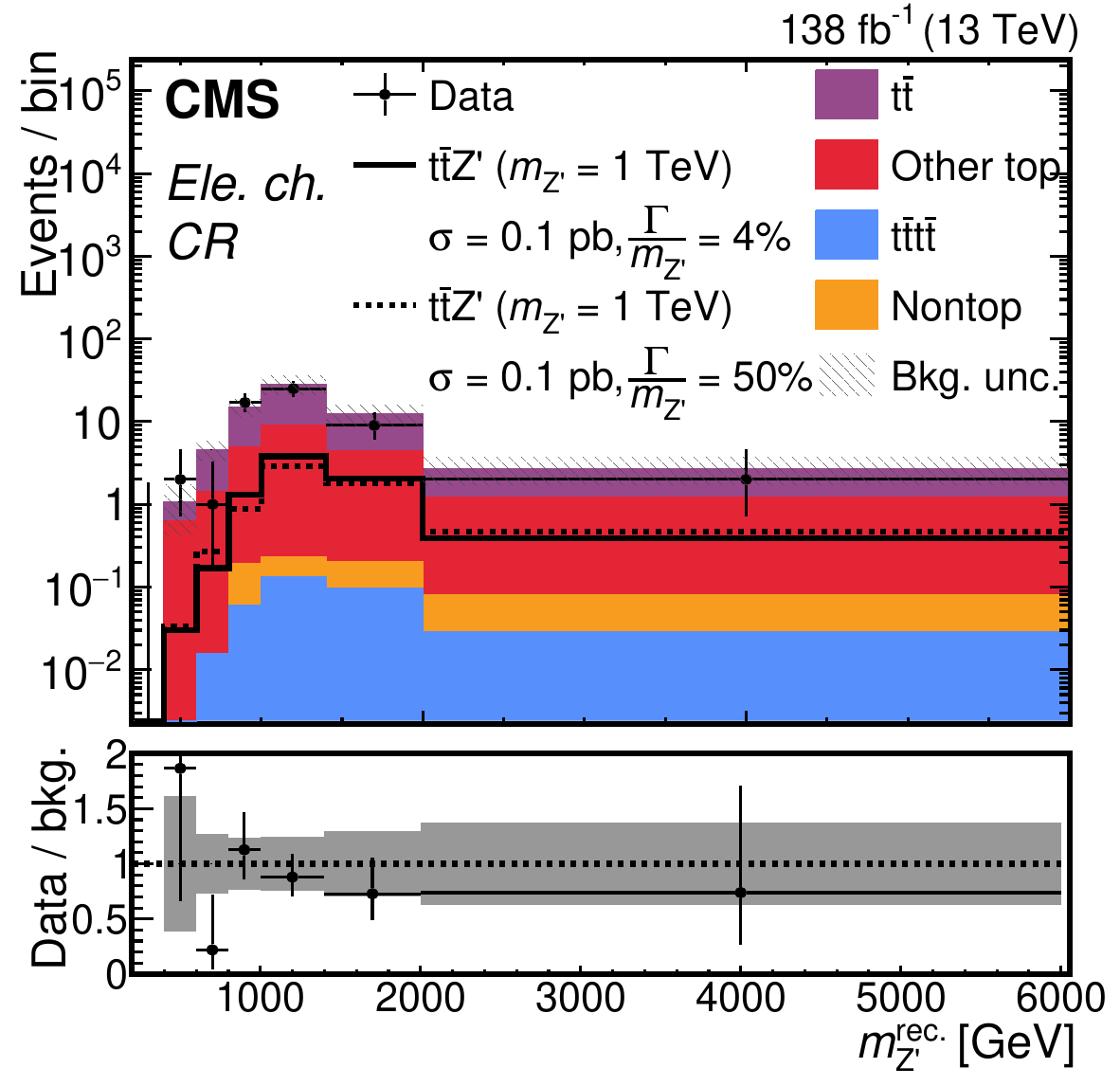}
\includegraphics[width=0.45\textwidth]{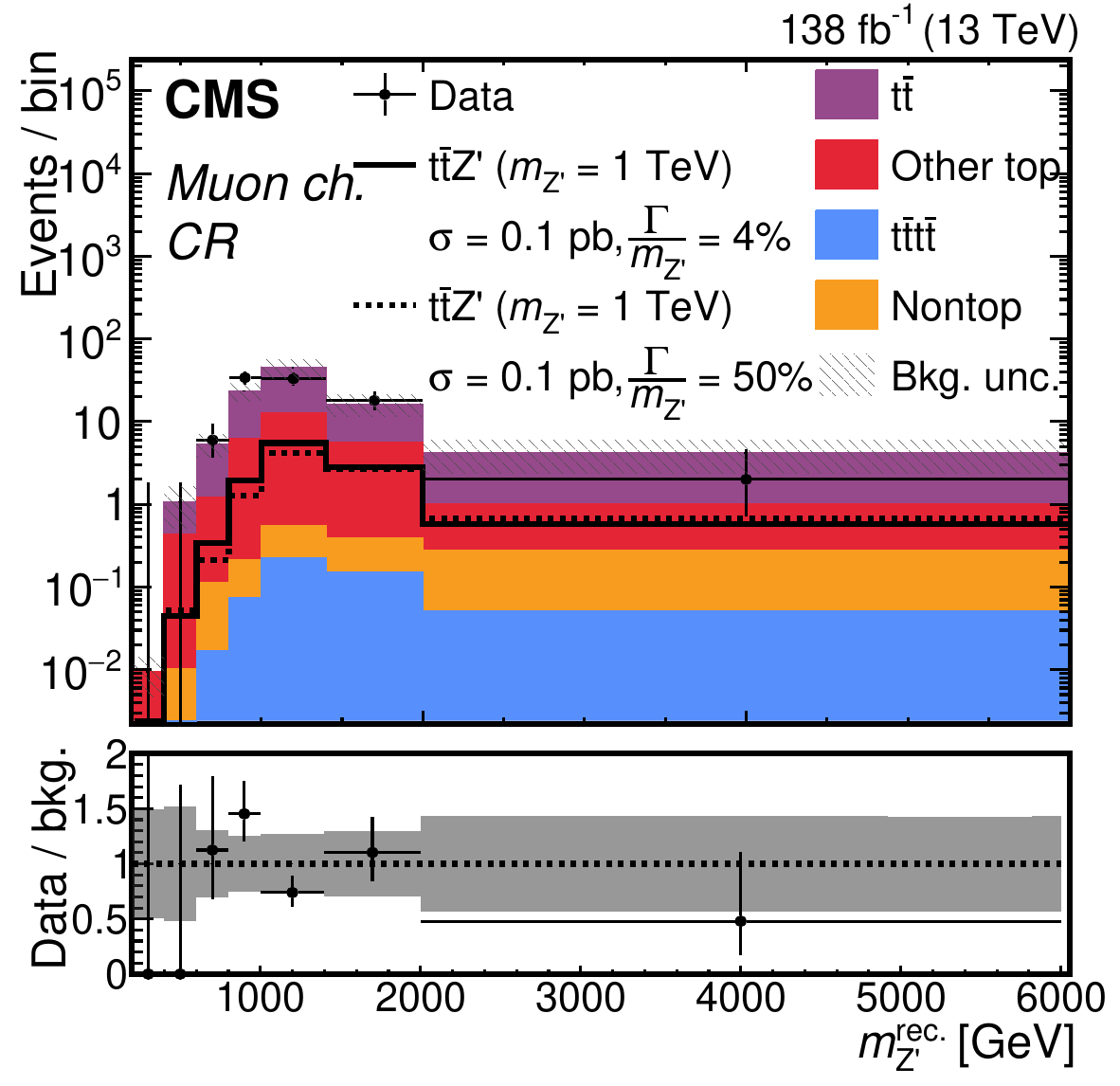}
\caption{Prefit distributions of \mzprec for the expected background (stacked histograms), data (black points), and two \zp signal hypotheses with a mass of 1\TeV and $\Gamma_{\zp}/m_{\zp} = 4$ (solid black line) or 50\% (dashed black line) for events in the electron (left) and muon (right) channels of the CR. Both signal hypotheses are normalized to a cross section of 0.1\pb. The hatched band in the upper panels represents the total uncertainty in the SM background expectation. The lower panels display the ratio of the data and the SM background, and the total uncertainty in the latter is shown as a grey band.}
\label{fig:cr_zpmass}
\end{figure}

Using the simulated samples, we have verified that the background composition remains consistent between the two regions, both in terms of the acceptance and efficiency of the events and in the shape of the \mzprec distribution for the dominant background processes. Figure~\ref{fig:cr_zpmass} shows the prefit \mzprec distributions for events selected in the CR for the muon and electron channels. The ratio between data and the SM background expectation is consistent with unity.

The background contribution in the SR is estimated from the simulated \mzprec distributions, with the sum of all background processes scaled by a correction factor to account for residual normalization differences between data and simulation. This correction factor is determined in the simultaneous maximum likelihood fit of the SR and CR, in which a freely floating parameter, correlated between the two regions, is employed to find the background normalization, allowing both the correction factor and the signal normalization to be derived. The normalization of each individual background process is also allowed to vary within the uncertainties described in Section~\ref{sec:systematics}.

\section{Systematic uncertainties}
\label{sec:systematics}

Systematic uncertainties from various sources are propagated to both the shape and normalization of the \mzprec distribution, affecting both signal and background. Shape uncertainties vary depending on the process, \zp boson mass, and data-taking year, so a range of values is quoted where necessary to reflect the minimum and maximum uncertainties observed under different conditions. Experimental uncertainties are considered uncorrelated between different years, as they primarily arise from statistical variations that are independent across the data-taking periods.

The trigger efficiency uncertainty range is 0.1--3\% for both background and signal samples. An additional correction is applied with an uncertainty of $\approx1\%$ to account for an inefficiency resulting from a shift in the timing of the electromagnetic calorimeter inputs to the first-level trigger in the region $\abs{\eta} > 2$ in the 2016 and 2017 data-taking periods~\cite{CMS:2020cmk}. The jet four-momenta are varied within the jet energy scale and resolution uncertainties, and are propagated to all affected variables including \ptmiss. These uncertainties are in the ranges 1.0--8.8 (1--34)\% for the top quark (nontop) background and 1--3\% for the signal. Top quark tagging uncertainties are in the ranges 5--27\% for the background and 10--39\% for the signal. The uncertainty in the background estimation is 10--20\%, as obtained from the simultaneous fit of the CR and SR described in Section~\ref{sec:Background_estimation}. These background normalizations from Section~\ref{sec:Background_estimation} are allowed to vary freely in the simultaneous fit of SR and CR and are treated as uncorrelated across the data-taking years.

The uncertainties in the tagging efficiency for heavy-flavor (\PQb and \PQc quarks) jets and the mistag rate for light-flavor (\PQu, \PQd, \PQs quarks) and gluon jets are propagated from uncertainties in the \PQb-tagging scale factors. These uncertainties vary within 0.3--4.9\% for the background and 0.9--4.1\% for the signal, and are considered to be partially correlated among the years of data taking. 

The integrated luminosities for the 2016, 2017, and 2018 data-taking years have 1.2--2.5\% individual uncertainties~\cite{CMS-LUM-17-003,CMS-PAS-LUM-17-004,CMS-PAS-LUM-18-002}, while the overall uncertainty for the 2016--2018 period is 1.6\%. The uncertainties related to the lepton reconstruction are below 2\% for all processes, and the uncertainty in the simulation of pileup is accounted for by varying the total inelastic $\Pp\Pp$ cross section by 4.6\%~\cite{CMS:2018mlc}, leading to uncertainties between 0.5 and 1.4\% for the background and 1.5\% for the signal. Both of these uncertainties are treated as correlated between the data-taking years.

Finally, the effect of the parton distribution function uncertainty on the acceptance is evaluated using the PDF4LHC recommendations~\cite{Butterworth:2015oua} and varies within 5--20\% for the background and 4--8\% for the signal. The systematic uncertainties associated with the renormalization and factorization scales are evaluated by varying each scale up and down by a factor of two, excluding the two extreme variations. This procedure results in an approximately constant uncertainty of about 40\% for the signal and 24--44\% for the background, reflecting the uncertainties within our phase space.

\section{Results}
\label{sec:results}

\begin{figure}[th]
\centering
\includegraphics[width=0.45\textwidth]{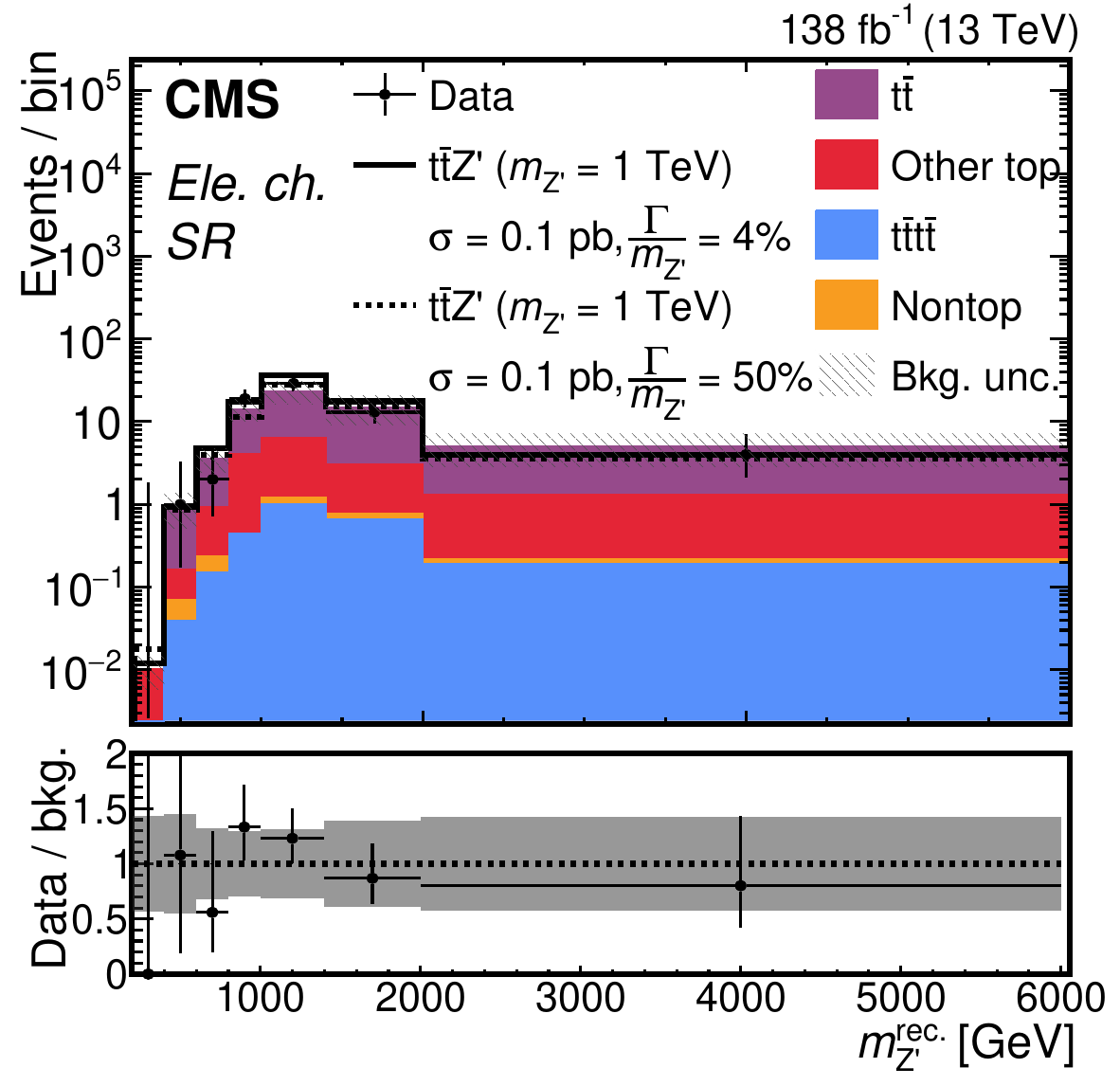}
\includegraphics[width=0.45\textwidth]{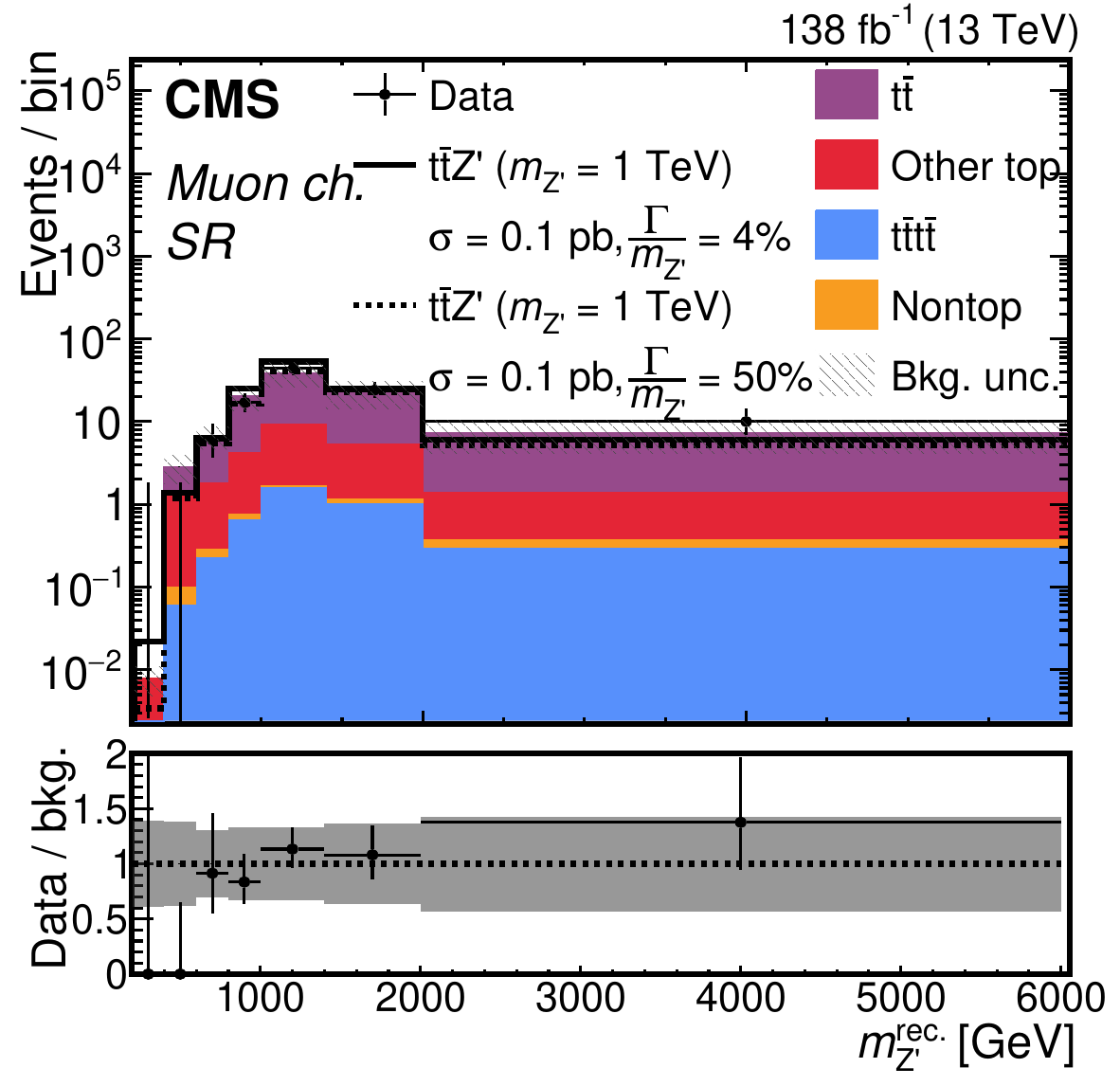}
\includegraphics[width=0.45\textwidth]{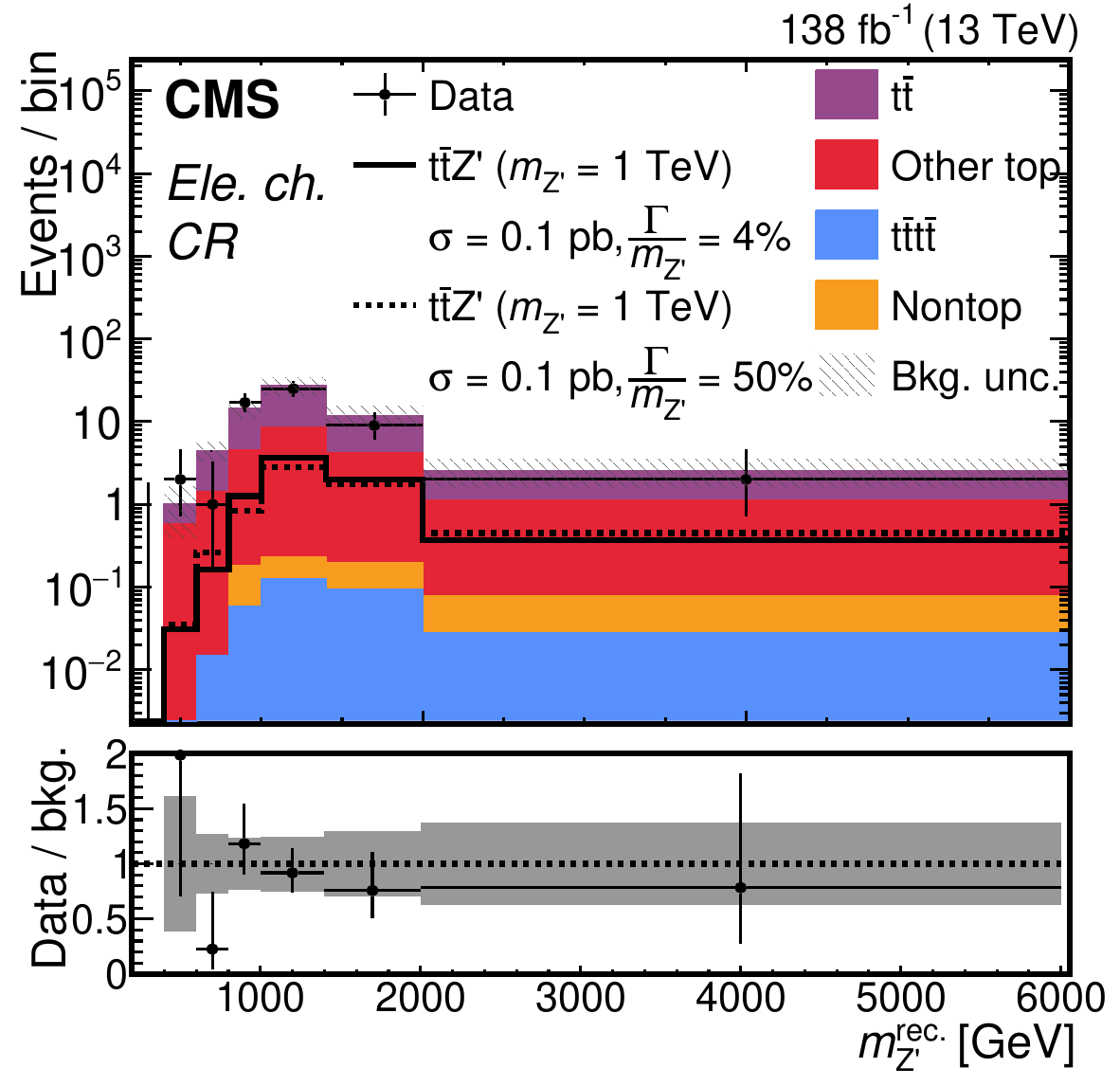}
\includegraphics[width=0.45\textwidth]{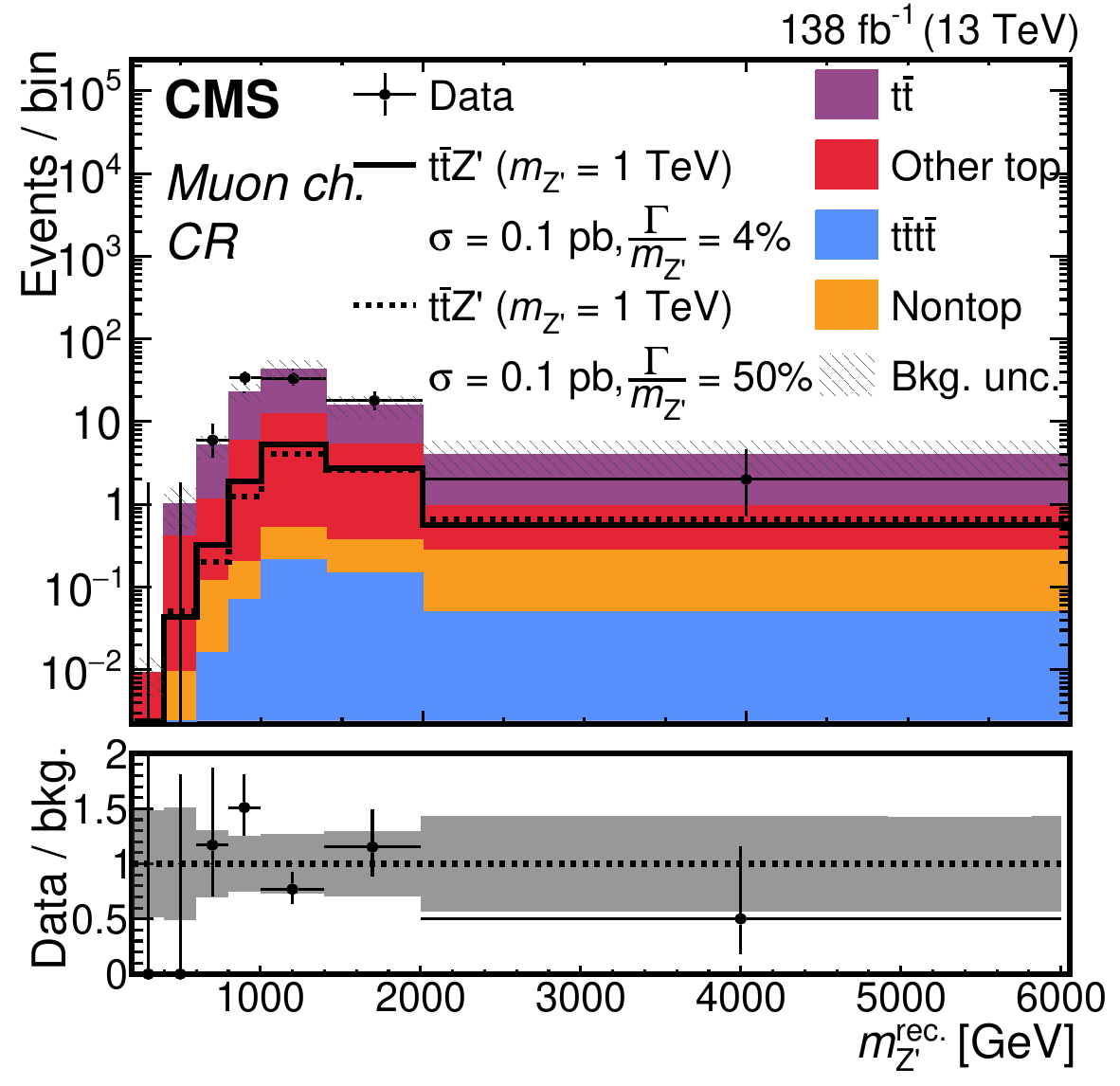}
\caption{Postfit distributions of \mzprec for the expected background (stacked histograms), data (black points), and two \zp signal hypotheses with a mass of 1\TeV and $\Gamma_{\zp}/m_{\zp} = 4$ (solid black line) or 50\% (dashed black line) for events in the electron (left) and muon (right) channels of the SR (upper row) and CR (lower row). Both signal hypotheses are normalized to a cross section of 0.1\pb. The hatched band in the upper panels represents the total uncertainty in the SM background expectation. The lower panels display the ratio of the data and the SM background, and the total uncertainty in the latter is shown as a grey band.}
\label{fig:zpmass_channel}
\end{figure}

The signal is extracted via a maximum likelihood fit of the expected signal and background to the data, performed simultaneously in the electron and muon channels of the SR and CR. The systematic uncertainties described in the previous section are included as nuisance parameters with log-normal probability distributions for normalization uncertainties and Gaussian distributions for shape uncertainties. Each bin of the fitted distributions is also assigned an uncertainty accounting for the limited number of simulated events, following the Barlow--Beeston "lite" approach~\cite{refBB1, refBB2}. The \mzprec distribution after a background-only fit to data is displayed in Fig.~\ref{fig:zpmass_channel}. The data agree with the SM background expectation within the uncertainties. We set upper limits at 95\% confidence level (\CL), using the asymptotic approximation, on the product of the cross section of the production of a \zp boson in association with a pair of top quarks, $\sigma(\Pp\Pp\to \ttbar\zp)$, and the branching fraction of its decay into two top quarks, $\mathcal{B}(\zp \to \ttbar)$. The following results have been determined using the $\CLs$ method~\cite{Junk:1999kv,Read:2002hq,Cowan:2010js} via the CMS statistical analysis tool \textsc{Combine}~\cite{CMS:2024onh}, which is based on the \textsc{RooFit}~\cite{verkerke2003roofittoolkitdatamodeling} and \textsc{RooStats}~\cite{moneta2011roostatsproject} frameworks.

\begin{figure}[ht]
\centering
\includegraphics[width=0.45\textwidth]{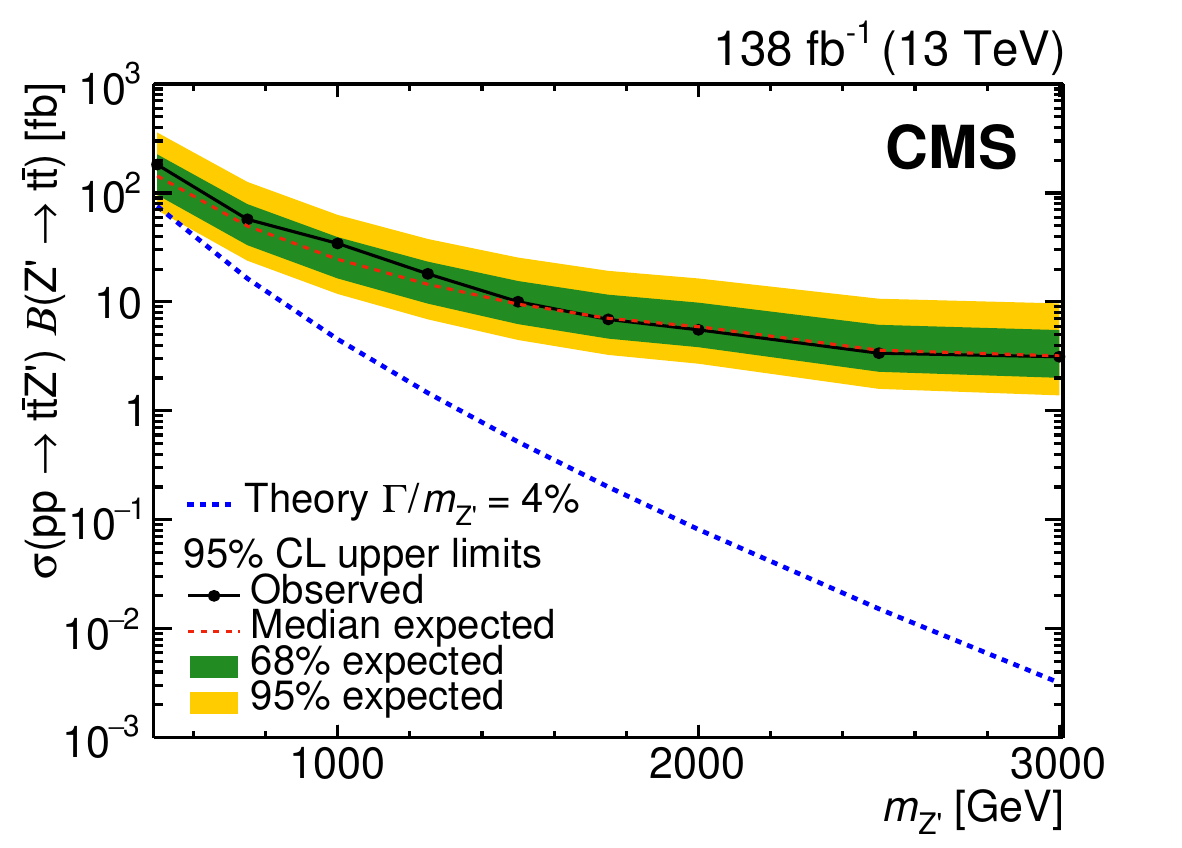}
\includegraphics[width=0.45\textwidth]{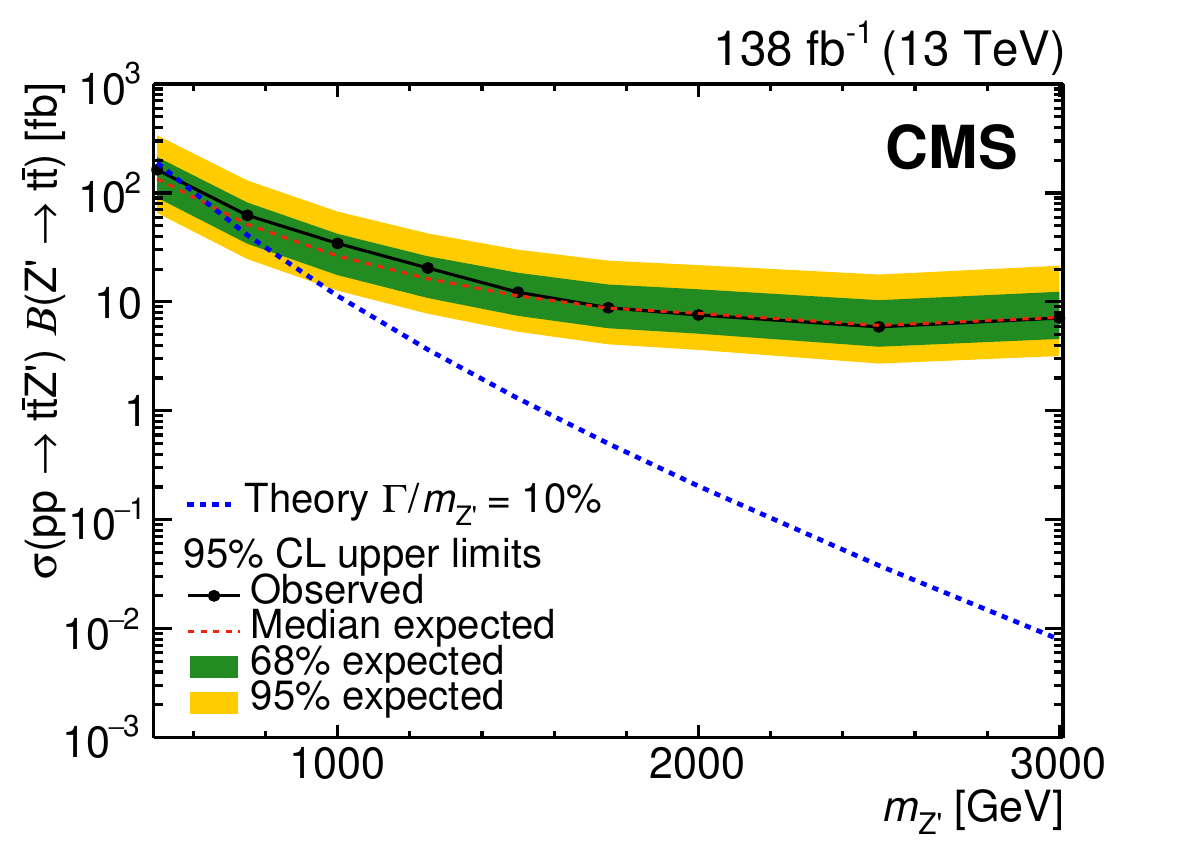}
\includegraphics[width=0.45\textwidth]{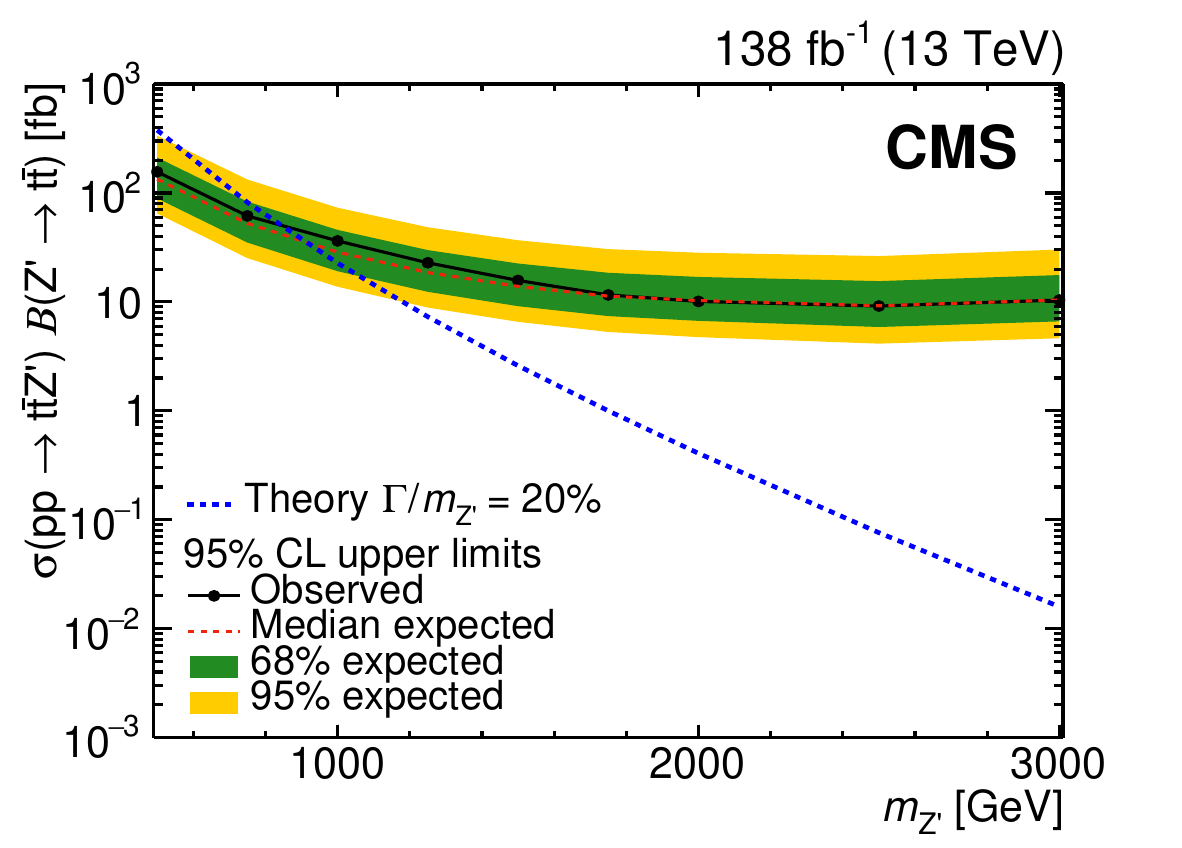}
\includegraphics[width=0.45\textwidth]{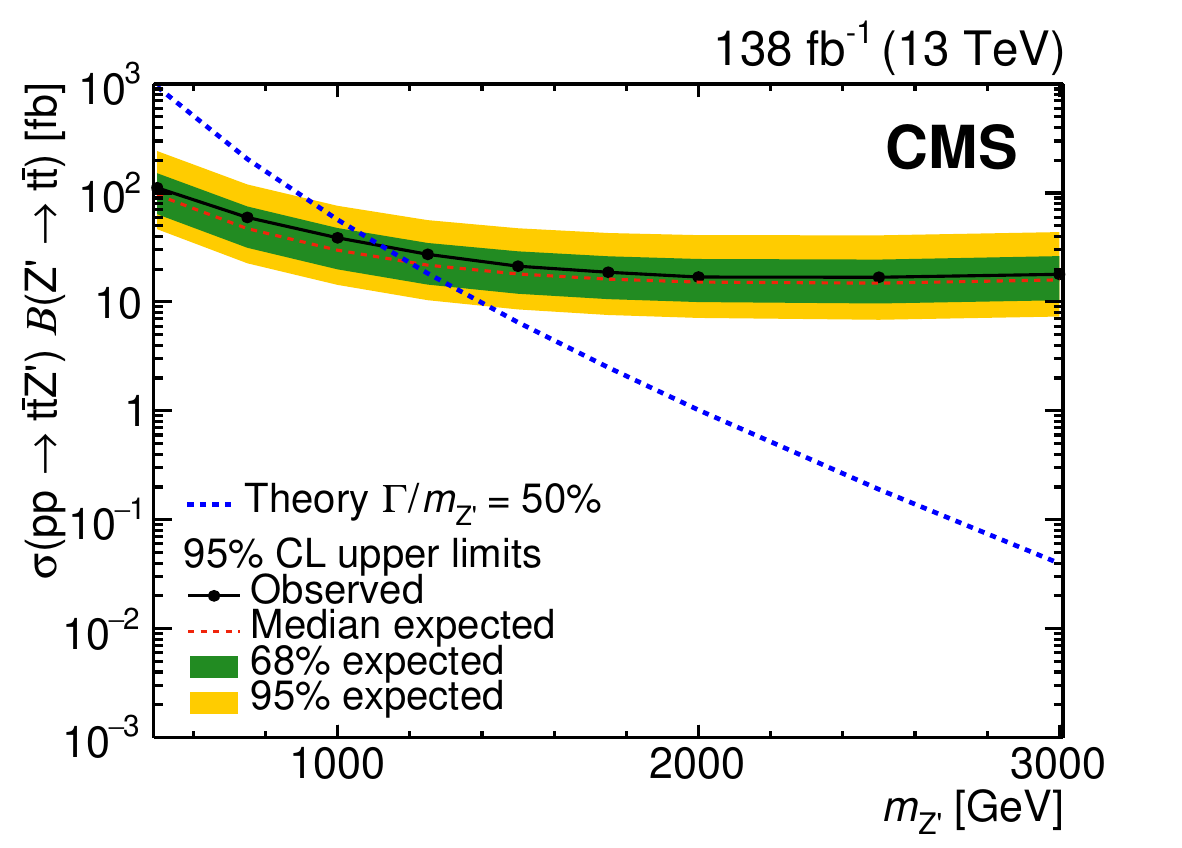}
\caption{Observed (black points) and expected (dashed red line) upper limits on $\sigma(\Pp\Pp\to \ttbar\zp)\mathcal{B}(\zp \to \ttbar)$ at 95\% CL as functions of the \zp boson mass for $\Gamma_{\zp}/m_{\zp} = 4$ (upper left), 10 (upper right), 20 (lower left), and 50\% (lower right), obtained combining the electron and muon channels and the 2016--2018 data-taking years. The inner (green) band and the outer (yellow) band indicate the regions containing 68 and 95\%, respectively, of the distribution of limits expected under the background-only hypothesis. The dashed blue line indicates the theoretical prediction at LO based on the model implementation in Ref.~\cite{Kim:2016plm}.}
\label{fig:Limit_width}
\end{figure}

The observed and expected upper limits on $\sigma(\Pp\Pp\to \ttbar\zp)\mathcal{B}(\zp \to \ttbar)$ as a function of the \zp boson mass are shown in Fig.~\ref{fig:Limit_width} for the combination of all data-taking periods and both electron and muon channels. The LO theoretical prediction of the $\sigma(\Pp\Pp \to \ttbar\zp)$ cross section, obtained using Ref.~\cite{Kim:2016plm}, is overlaid. The observed upper limits range from 170 to 3\fb for $\Gamma_{\zp}/m_{\zp} = 4\%$ in the mass range of 500 to 3000\GeV. For values of $\Gamma_{\zp}/m_{\zp}$ of 10, 20, and 50\% the ranges are 160--6, 160--9, and 110--17\fb, respectively. These correspond to lower limits on the \zp boson mass of 560, 850, and 1130\GeV for values of $\Gamma_{\zp}/m_{\zp}$ of 10, 20, and 50\%, respectively, which are the most stringent limits to date. 

\clearpage

\section{Summary}
\label{sec:summary}

A search for a top-philic \zp boson produced in association with a pair of top quarks and decaying into a top quark pair has been presented. It is based on a sample of proton-proton collision data collected at $\sqrt{s} = 13$\TeV, by the CMS experiment at the CERN LHC in 2016--2018, corresponding to an integrated luminosity of $\runtwolumi\fbinv$. The search focuses on the single-lepton (electron or muon) channel, in association with hadronically decaying top quarks. The case of a significant mass difference between the \zp boson and the top quark is considered, resulting in highly Lorentz-boosted top quarks reconstructed as large-radius jets using the \textsc{ParticleNet} algorithm, which uses advanced machine-learning techniques to improve the signal-to-background ratio and enhances the search sensitivity. The reconstructed \zp boson mass is used to probe for the presence of a beyond-the-standard-model signal. The observations are consistent with standard model predictions. Exclusion limits are set on a top-philic \zp boson as a function of the \zp boson mass and for widths ranging from 4 to 50\% of its mass. The lower limits on the $\ttbar\zp$ production cross section, at 95\% confidence level, range between 170 and 3\fb for a relative width of 4\% over the mass range between 500 and 3000\GeV. For widths of 10, 20, and 50\%, the ranges are 160--6, 160--9, and 110--17\fb, respectively, corresponding to lower limits on the \zp boson mass of 560, 850, and 1130\GeV, respectively. These results represent the most stringent limits to date on the existence of a \zp boson that couples exclusively to top quarks.

\begin{acknowledgments}
We congratulate our colleagues in the CERN accelerator departments for the excellent performance of the LHC and thank the technical and administrative staffs at CERN and at other CMS institutes for their contributions to the success of the CMS effort. In addition, we gratefully acknowledge the computing centres and personnel of the Worldwide LHC Computing Grid and other centres for delivering so effectively the computing infrastructure essential to our analyses. Finally, we acknowledge the enduring support for the construction and operation of the LHC, the CMS detector, and the supporting computing infrastructure provided by the following funding agencies: SC (Armenia), BMFWF and FWF (Austria); FNRS and FWO (Belgium); CNPq, CAPES, FAPERJ, FAPERGS, and FAPESP (Brazil); MES and BNSF (Bulgaria); CERN; CAS, MoST, and NSFC (China); MINCIENCIAS (Colombia); MSES and CSF (Croatia); RIF (Cyprus); SENESCYT (Ecuador); ERC PRG and PSG, TARISTU24-TK10 and MoER TK202 (Estonia); Academy of Finland, MEC, and HIP (Finland); CEA and CNRS/IN2P3 (France); SRNSF (Georgia); BMFTR, DFG, and HGF (Germany); GSRI (Greece); MATE and NKFIH (Hungary); DAE and DST (India); IPM (Iran); SFI (Ireland); INFN (Italy); MSIT and NRF (Republic of Korea); MES (Latvia); LMTLT (Lithuania); MOE and UM (Malaysia); BUAP, CINVESTAV, CONACYT, LNS, SEP, and UASLP-FAI (Mexico); MOS (Montenegro); MBIE (New Zealand); PAEC (Pakistan); MSHE, NSC, and NAWA (Poland); FCT (Portugal);  MESTD (Serbia); MICIU/AEI and PCTI (Spain); MOSTR (Sri Lanka); Swiss Funding Agencies (Switzerland); MST (Taipei); MHESI (Thailand); TUBITAK and TENMAK (T\"{u}rkiye); NASU (Ukraine); STFC (United Kingdom); DOE and NSF (USA).

\begin{sloppypar}
\setlength\emergencystretch{\hsize}
\hyphenation{Rachada-pisek} Individuals have received support from the Marie-Curie programme and the European Research Council and Horizon 2020 Grant, contract Nos.\ 675440, 724704, 752730, 758316, 765710, 824093, 101115353, 101002207, 101001205, and COST Action CA16108 (European Union); the Leventis Foundation; the Alfred P.\ Sloan Foundation; the Alexander von Humboldt Foundation; the Science Committee, project no. 22rl-037 (Armenia); the Fonds pour la Formation \`a la Recherche dans l'Industrie et dans l'Agriculture (FRIA) and Fonds voor Wetenschappelijk Onderzoek contract No. 1228724N (Belgium); the Beijing Municipal Science \& Technology Commission, No. Z191100007219010, the Fundamental Research Funds for the Central Universities, the Ministry of Science and Technology of China under Grant No. 2023YFA1605804, the Natural Science Foundation of China under Grant No. 12535004, and USTC Research Funds of the Double First-Class Initiative No.\ YD2030002017 (China); the Ministry of Education, Youth and Sports (MEYS) of the Czech Republic; the Shota Rustaveli National Science Foundation (Georgia); the Deutsche Forschungsgemeinschaft (DFG), among others, under Germany's Excellence Strategy -- EXC 2121 ``Quantum Universe" -- 390833306, and under project number 400140256 - GRK2497; the Hellenic Foundation for Research and Innovation (HFRI), Project Number 2288 (Greece); the Hungarian Academy of Sciences, the New National Excellence Program - \'UNKP, the NKFIH research grants K 131991, K 138136, K 143460, K 143477, K 147557, K 146913, K 146914, K 147048, TKP2021-NKTA-64, and 2025-1.1.5-NEMZ\_KI-2025-00004, and MATE KKP and KKPCs Research Excellence and Flagship Research Groups grants (Hungary); the Council of Science and Industrial Research, India; ICSC -- National Research Centre for High Performance Computing, Big Data and Quantum Computing, FAIR -- Future Artificial Intelligence Research, and CUP I53D23001070006 (Mission 4 Component 1), funded by the NextGenerationEU program, the Italian Ministry of University and Research (MUR) under Bando PRIN 2022 -- CUP I53C24002390006, PRIN PRIMULA 2022RBYK7T (Italy); the Latvian Council of Science; the Ministry of Science and Higher Education, project no. 2022/WK/14, and the National Science Centre, contracts Opus 2021/41/B/ST2/01369, 2021/43/B/ST2/01552, 2023/49/B/ST2/03273, and the NAWA contract BPN/PPO/2021/1/00011 (Poland); the Funda\c{c}\~ao para a Ci\^encia e a Tecnologia (Portugal); the National Priorities Research Program by Qatar National Research Fund; MICIU/AEI/10.13039/501100011033, ERDF/EU, ``European Union NextGenerationEU/PRTR", projects PID2022-142604OB-C21, PID2022-139519OB-C21, PID2023-147706NB-I00, PID2023-148896NB-I00, PID2023-146983NB-I00, PID2023-147115NB-I00, PID2023-148418NB-C41, PID2023-148418NB-C42, PID2023-148418NB-C43, PID2023-148418NB-C44, PID2024-158190NB-C22, RYC2021-033305-I, RYC2024-048719-I, CNS2023-144781, CNS2024-154769 and Plan de Ciencia, Tecnolog{\'i}a e Innovaci{\'o}n de Asturias, Spain; the Chulalongkorn Academic into Its 2nd Century Project Advancement Project, the National Science, Research and Innovation Fund program IND\_FF\_68\_369\_2300\_097, and the Program Management Unit for Human Resources \& Institutional Development, Research and Innovation, grant B39G680009 (Thailand); the Eric \& Wendy Schmidt Fund for Strategic Innovation through the CERN Next Generation Triggers project under grant agreement number SIF-2023-004; the Kavli Foundation; the Nvidia Corporation; the SuperMicro Corporation; the Welch Foundation, contract C-1845; and the Weston Havens Foundation (USA).
\end{sloppypar}
\end{acknowledgments}\section*{Data availability} Release and preservation of data used by the CMS Collaboration as the basis for publications is guided by the  \href{https://doi.org/10.7483/OPENDATA.CMS.1BNU.8V1W}{CMS data preservation, re-use and open access policy}.

\bibliography{auto_generated}

\cleardoublepage \appendix\section{The CMS Collaboration \label{app:collab}}\begin{sloppypar}\hyphenpenalty=5000\widowpenalty=500\clubpenalty=5000\cmsinstitute{Yerevan Physics Institute, Yerevan, Armenia}
{\tolerance=6000
A.~Hayrapetyan, V.~Makarenko\cmsorcid{0000-0002-8406-8605}, A.~Tumasyan\cmsAuthorMark{1}\cmsorcid{0009-0000-0684-6742}
\par}
\cmsinstitute{Institut f\"{u}r Hochenergiephysik, Vienna, Austria}
{\tolerance=6000
W.~Adam\cmsorcid{0000-0001-9099-4341}, L.~Benato\cmsorcid{0000-0001-5135-7489}, T.~Bergauer\cmsorcid{0000-0002-5786-0293}, M.~Dragicevic\cmsorcid{0000-0003-1967-6783}, P.S.~Hussain\cmsorcid{0000-0002-4825-5278}, M.~Jeitler\cmsAuthorMark{2}\cmsorcid{0000-0002-5141-9560}, N.~Krammer\cmsorcid{0000-0002-0548-0985}, A.~Li\cmsorcid{0000-0002-4547-116X}, D.~Liko\cmsorcid{0000-0002-3380-473X}, M.~Matthewman, J.~Schieck\cmsAuthorMark{2}\cmsorcid{0000-0002-1058-8093}, R.~Sch\"{o}fbeck\cmsAuthorMark{2}\cmsorcid{0000-0002-2332-8784}, M.~Shooshtari\cmsorcid{0009-0004-8882-4887}, M.~Sonawane\cmsorcid{0000-0003-0510-7010}, W.~Waltenberger\cmsorcid{0000-0002-6215-7228}, C.E.~Wulz\cmsAuthorMark{2}\cmsorcid{0000-0001-9226-5812}
% ...
\par}
$^{1}$Also at Yerevan State University, Yerevan, Armenia\\
$^{2}$Also at Technische Universit\"{a}t Wien, Vienna, Austria\\
% ...
\end{sloppypar}
\end{document}